\documentclass[12pt,preprint]{aastex}

\begin{document}

\title{Hubble Space Telescope observations of an extraordinary flare in the M87 jet\altaffilmark{1}}


\author{
Juan P. Madrid \altaffilmark{2,3}     
}

\altaffiltext{1}{Based on  observations made with  the NASA/ESA Hubble
  Space Telescope, obtained at  the Space Telescope Science Institute,
  which is operated by the Association of Universities for Research in
  Astronomy,   Inc.,   under  NASA   contract   NAS  5-26555.    These
  observations are associated with programs 9474, 10133, 10617.}

\altaffiltext{2}{Space  Telescope Science Institute,  3700 San Martin
Drive, Baltimore, MD 21218}

\altaffiltext{3}{ Now at Department of Physics and Astronomy, McMaster
  University, Hamilton, ON, L8S 4M1, Canada}


\begin{abstract}

  HST-1,  a knot  along the  M87  jet located  $0.85\arcsec$ from  the
  nucleus  of  the  galaxy  has experienced  dramatic  and  unexpected
  flaring  activity since early  2000. We  present analysis  of Hubble
  Space Telescope  Near-Ultraviolet (NUV) imaging of the  M87 jet from
  1999 May  to 2006  December that reveals  that the NUV  intensity of
  HST-1 has increased 90 times  over its quiescent level and outshines
  the  core of  the galaxy.   The NUV  light curve  that we  derive is
  synchronous with  the light curves derived in  other wavebands.  The
  correlation of  X-ray and  NUV light curves  during the  HST-1 flare
  confirms the  synchrotron origin  of the X-ray  emission in  the M87
  jet.   The outburst observed  in HST-1  is at  odds with  the common
  definition  of  AGN  variability   usually  linked  to  blazars  and
  originating in close  proximity of the central black  hole. In fact,
  the  M87 jet is  not aligned  with our  line of  sight and  HST-1 is
  located  at one  million Schwarzchild  radii from  the super-massive
  black hole in the core of the galaxy.

\end{abstract}

\keywords{galaxies: jets - galaxies: active}


\section{Introduction} 
                            
M87, the cD  galaxy of the Virgo cluster, is  a giant elliptical famed
for its spectacular galactic-scale  plasma jet.  Due to its proximity,
images of  M87's jet at high  resolution have revealed  a profusion of
distinct knots or regions of  enhanced emission along the whole length
of the jet. These knots  have been clearly detected at radio, optical,
UV,  and X-ray  wavelengths.   The  detection of  these  UV and  X-ray
emission regions  hundreds of  parsecs away from  the AGN  proves that
these  are regions  of in  situ particle  acceleration within  the jet
because  such high  energy  emission vanishes  rapidly. The  radiative
half-lives of synchrotron X-ray emitting electrons are of the order of
years, and the cooling time for UV emitting particles are of the order
of decades (Harris \& Krawczynski 2006). High energy emission would be
confined in a small space  without re-acceleration along the jet. Thus
these knots must be regions of acceleration distinct from the AGN.

Until February  2000 HST-1  was an inconspicuous  knot of the  M87 jet
located $0.85\arcsec$ from  the nucleus of the galaxy  (Waters \& Zepf
2005).   Since  that  date,  HST-1  has shown  an  unexpectedly  rapid
variability in  all wavebands. More  strikingly, in 2003  HST-1 became
brighter than  the nucleus of the  galaxy, which is known  to harbor a
super-massive    black     hole    of    $3.2     \pm    0.9\times10^9
M_{\odot}$(Macchetto et  al. 1997).  HST-1  is also the  most probable
site  of  production of  the  TeV  $\gamma$  rays emanating  from  M87
recently reported  by the HESS  collaboration (Aharonian et  al. 2006,
see also Cheung et al. 2007).

Specific  observations  of HST-1  have  been  carried  out across  the
electromagnetic  spectrum, and  a particularly  detailed study  of the
flaring of HST-1 has been conducted with the Chandra X-ray Observatory
by D. E.  Harris and  collaborators (Harris et al.  2003, 2006, 2008).
The X-ray intensity  of this peculiar knot has  increased more than 50
times in the past five years and  peaked in 2005. There is a wealth of
high-quality HST  NUV imaging data of  the M87 jet that  has been only
succinctly  presented in  the  past  (Madrid et  al.   2007). The  NUV
light-curve for HST-1 that we  present here has broadly the same shape
as the X-ray light curve presented  by Harris et al. (2006): the flare
rises, peaks, and declines simultaneously in the X-rays and the NUV.

HST-1  is located  at one  million  Schwarzchild radii  away from  the
galactic nucleus  but if  M87 were  at a greater  distance, or  if our
telescopes  had   lesser  resolution,  this  flare   would  have  been
interpreted as variability intrinsic to the central black hole and its
immediate vicinity. This blazar-like behavior is clearly isolated from
the central  engine and it  is not directly  beamed as the M87  jet is
misaligned  with respect to  the line  of sight  (Harris et  al. 2006,
Cheung et al. 2007). A detailed characterization of this flare is thus
important to better understand blazar variability.

We describe the Hubble Space  Telescope view of the remarkable flaring
of the  HST-1 knot with high  resolution imaging taken  over more than
seven years of  observations and aim to present  the visually striking
NUV data that bridges the gap  between the X-ray (Harris et al.  2006)
and radio (Cheung et al.  2007) observations of HST-1.

\section{Observations \& Data Reduction} 

We present observations obtained with two instruments on board HST: the
Space Telescope  Imaging Spectrograph  (STIS) and the  Advanced Camera
for Surveys (ACS).

STIS stopped functioning in 2004  August due to an electronics failure
on the redundant (Side 2)  power supply system. All observations after
August 2004  were taken with  the ACS. Even  though each of  these two
instruments has unique characteristics,  they cover the same wavebands
and provide data  that are easily compared.  Moreover,  the ACS images
have the same file structure  as STIS images making the data reduction
procedure very  similar between  the two instruments.  The discrepancy
between the STIS and the ACS absolute photometric calibration does not
exceed 2\% (Bohlin, 2007).

The  STIS observations were  carried out  using the  NUV/MAMA detector
which  has a  field of  view of  $24.7\arcsec\times24.7\arcsec$  and a
$0.024\arcsec$pixel  size. The  M87  jet was  imaged  with the  F25QTZ
filter    that   has    its   maximum    throughput    wavelength   at
$2364.8\mbox{\AA}$ and a width of $995.1\mbox{\AA}$. Due to the nature
of the detector NUV/MAMA images  are free of cosmic rays (Kim Quijano,
2003).

The ACS High Resolution Camera (HRC)  is a CCD instrument with a field
of    view   of    $29\arcsec\times25\arcsec$   and    a    scale   of
$\sim0.025\arcsec$  per pixel.   We analyze  images acquired  with the
F220W and F250W filters, which  are the two broadband NUV filters with
the most similar characteristics to  the STIS F25QTZ filter. The F220W
filter has its  pivot wavelength at $2255.5\mbox{\AA}$ and  a width of
$187.3\mbox{\AA}$, for  the F250W these  values are $2715.9\mbox{\AA}$
and $239.4\mbox{\AA}$  respectively (Mack et al. 2003,  Gonzaga et al.
2005).

We obtained  the flatfielded science  files (FLT) from the  HST public
archive for  data acquired by  both instruments.  These  science ready
files are  processed through  the automatic reduction  and calibration
pipeline (CALACS)  before they are retrieved from  the public archive.
The  pipeline subtracts  the bias  and  dark current  and applies  the
flatfield to the raw CCD data (Sirianni et al. 2005).  Subsequent data
reduction  was performed  using the  software package  Space Telescope
Science Data Analysis System (STSDAS).

We analyzed data taken over a period of time of more than seven years,
from 1999 May,  through 2006 December.  Each image,  at all epochs, is
the product of  the combination of four single  exposures taken within
the same  orbit.  This allows us  to eliminate cosmic rays  in the ACS
images and improve the signal to noise.

We  used the  STSDAS task  {\sc multidrizzle}  to apply  the geometric
distortion correction,  eliminate cosmic  rays, and align  and combine
the individual  exposures of  every epoch.  The  distortion correction
was computed  with up-to-date distortion  coefficient tables retrieved
from the  Multimission Archive at the Space  Telescope (MAST).  During
the data  reduction process we  preserved the native pixel  size.  The
final  output images  generated by  {\sc multidrizzle}  have  units of
counts per second for the STIS data, and units of electrons per second
for the ACS data (Koekemoer et al. 2002).

We derive  fluxes and  errors by doing  aperture photometry  with {\sc
  phot} with and aperture radius of 10 pixels or $0.25\arcsec$. At the
distance of M87, 16.1 Mpc  (Tonry et al. 2001), $1\arcsec$ corresponds
to 77pc. We convert the number of counts obtained with {\sc phot} into
flux and flux errors by  using {\sc photflam}, or inverse sensitivity,
for  each  instrument  found  in  the  updated  ACS  zeropoint  tables
maintained by the STScI, or in the header of the STIS images:

\begin{mathletters}
{\sc photflam$_{STIS/F25QTZ}$}=5.8455.10$^{-18}$ erg cm$^{-2}$ \AA\\
\end{mathletters}

\begin{mathletters}
{\sc photflam$_{ACS/F220W}$}=8.0721.10$^{-18}$ erg cm$^{-2}$ \AA\\
\end{mathletters}

\begin{mathletters}
{\sc photflam$_{ACS/F250W}$}=4.7564.10$^{-18}$ erg cm$^{-2}$ \AA\\
\end{mathletters}

Once the  calibrated fluxes  and poisson errors  are derived  they are
transformed into  miliJanskys using the  pyraf task {\sc  calcphot} of
the  synthetic  photometry  ({\sc  synphot})  software  package  under
STSDAS.  We  also scale the flux measurements  obtained with different
bands using {\sc  synphot} (Laidler et al.  2005).   We expect that no
additional  errors are  introduced  by {\sc  synphot}  when doing  the
transformation to miliJanskys. We assume that the spectrum of HST-1 is
described by a  power law with index $\alpha$  (Perlman et al.  2001),
and we define the flux density as $S_{\nu}\propto\nu^{-\alpha}$.

The background light was estimated by measuring the flux with the same
circular aperture  at the same  radial distance from the  nucleus than
HST-1,  but on  the side  of the  jet. The  aperture  corrections were
performed using  the values  published by Proffitt  et al.  (2003) for
STIS and  Sirianni et al.  (2005)  for the ACS.  We  use the reddening
towards M87  determined by Schlegel  et al. (1998),  E(B-V)=0.022, and
the extinction  relations from Cardelli  et al.  (1989) to  derive the
extinction in the  HST filters.  We find the  following values for the
extinction:       $A_{F25QTZ}=0.190$,      $A_{F220W}=0.220$,      and
$A_{F250W}=0.134$.


\section{Results} 

HST-1 was dormant until 2000  February when its flaring activity began
(Waters \& Zepf 2003).  We see this in Figure 1 which is a zoom of the
inner regions  of the  M87 jet  and displays, on  the left,  the early
evolution  of HST-1.   Three main  emission loci  are visible  in this
zoom, from left to right, these are: the nucleus of the galaxy, HST-1,
and knot D. The images in the left column of Fig. 1 were acquired with
STIS/F25QTZ while the  images on the right column  were taken with the
ACS in the F220W band.

In the  STIS image of  1999 May (top  left) HST-1 was  an unremarkable
knot  along  the  M87  jet.   The  brightening  of  HST-1  is  already
noticeable in 2001  July.  The images in the lower  left were taken in
2002 February and 2002 July respectively and show the slow brightening
of  HST-1 during this  year.  At  the end  of 2002  HST-1 is  15 times
brighter than in 1999 May.

In 2003 HST-1 became dramatically variable. The image on the top right
was taken on 2003 April as  HST-1 continues to rise in flux.  HST-1 is
at  its highest  recorded  brightness in  2005  May, on  this date  we
recorded the  highest flux of HST-1  in the NUV, namely  0.54 mJy.  At
this point in  time the NUV flux of HST-1 was  four times the measured
flux of the central engine of  the galaxy.  The peak of the X-ray flux
based on  Chandra observations was  reported on 2005 April  (Harris et
al.  2006). HST-1 attained a NUV  flux 90 times its quiescent level in
2005 May.   The HST data  acquired in 1999  gives us a measure  of the
brightness of  HST-1 during its latent  state and allow  us to measure
the total factor by which  the brightness changed during the outburst.
Chandra was  just launched  in 1999 and  VLBI radio  observations date
back to 2000 only (Cheung et al. 2007).

After May 2005  HST-1 declined in intensity with  a decay time similar
to  the rise  time.  HST-1  experienced a  second and  also unexpected
outburst in  2006 November. This  second outburst is fainter  than the
first one in 2005  May. The image on the lower right  of Fig.  1 shows
HST-1 during the second, yet fainter, outburst in 2006 November. It is
evident from Figure 1 that the jet itself is better mapped by the STIS
images.

Table 1 present the log of observations of the M87 jet taken with STIS
and ACS.   This table  also presents the  NUV intensities  and poisson
errors of  the nucleus of the  galaxy and HST-1 at  all epochs studied
here.   Although   magnitudes  of   the  Space  Telescope   system  or
erg/s/cm$^2$/Hz would  be more  natural units we  decided to  plot our
light curve in mJy to facilitate comparison with observations at other
wavebands (Waters  \& Zepf 2005, Perlman  et al.  2003,  Harris et al.
2006).

Figure 2 shows the light curve of HST-1 and the nucleus of the galaxy.
The HST-1 light curve is bumpy in the radio and the X-rays and the NUV
is  not an exception.  The dramatic  flaring of  HST-1 can  be clearly
appreciated in this figure.

Table 2 contains  the doubling and halving times  for HST-1 calculated
for the  NUV following  the prescriptions of  Harris et al  (2006). We
calculate  $y=I_{2}/I_{1}$,  the flux  ratio,  and  the time  elapsed,
$\Delta$t, between two consecutive observations.  The doubling time is
calculated using  DT=$[\frac{1}{y-1}]\Delta$t and the  halving time by
HT=$[\frac{0.5}{1-y}]\Delta$t. The bumpiness attributed to synchrotron
losses in  the X-ray  persists in the  NUV. These rapid  variations of
brightness  are  consistent  with  the month  time-scale  variability
reported by Perlman  et al.  (2003) for the early  stages of the flare
and found here to persist through time.

The X-ray and NUV light curves of HST-1 are plotted together in Figure
3.   We performed  a formal  correlation analysis  of these  two light
curves  by taking  thirty simultaneous  values  of the  NUV and  X-ray
fluxes and deriving the  Spearman rank correlation coefficient $\rho$.
This coefficient  is a non  parametric measure of correlation  and the
range is  $0<\rho<1$, the higher  the value, the more  significant the
correlation (Wall \& Jenkins  2003).  Simultaneous measurements of the
flux of HST-1 in the X-ray and the NUV yield $\rho=0.966$ reflecting a
strong correlation.


\section{Discussion} 


In Harris et al.  (2003) synchrotron loss models based on the NUV data
available at that  time predicted the optical decay  time of the flare
to be a factor  of 10 larger than the X-ray decay  time.  On the other
hand, Perlman et al. (2003)  forethought a similar decay timescale for
both   optical  and   X-ray  lighcurves. 

The  simultaneous  rise  and  fall  of  the flare  at  NUV  and  X-ray
wavelengths supports  the first  plausible hypothesis of  the physical
origin of the HST-1 flare postulated by Harris et al.  (2006), namely,
that  a simple  compression  caused the  HST-1 outburst.   Compression
increases both the magnetic field  strength and particle energy at all
wavelengths equally, leading to simultaneous flaring at all wavebands.
The  magnetic field  vectors in  HST-1  are perpendicular  to the  jet
direction also consistent with a shock (Perlman et al. 2003).

The overall rise  and fall timescales, similar in  both bands, and the
lack of  a large delay  between bands suggest  a rapid expansion  as a
probable  cause  for the  decrease  in  luminosity.   However, a  more
rigorous analysis of the rise and fall timescales shows that expansion
is  not the dominant  mechanism of  energy loss  for HST-1,  see below
(Harris et al. 2008).

A more  elaborated theoretical interpretation for the  origin of HST-1
was  presented  by  Stawarz  et  al.   (2006)  and  supported  by  the
observations  of Cheung  et al.(2007).   This newer  hypothesis claims
that HST-1 originates  in a nozzle throat of the  M87 jet that creates
reconfinement  of magnetic  field  lines liberating  large amounts  of
energy,  similar to  the process  responsible for  solar  flares.  The
gravitational influence of the  central AGN on the velocity dispersion
of the  stars in the  innermost regions of  this galaxy has  been well
documented with early  HST observations (Lauer et al.   1992, see also
Macchetto et al. 1997).  The hot thermal gas can be expected to follow
the  distribution of  the  stars in  this  inner region  and create  a
reconfinement shock  in the jet  due to an enhanced  thermal pressure.
This reconfinement should  happen at roughly the same  distance of the
well known  stellar cusp, precisely  where HST-1 lays (Stawarz  et al.
2006).

The  doubling and halving  time scales  of the  NUV presented  in this
paper and  the X-rays ones published  by Harris et al.   (2006) do not
always  perfectly overlap in  time, the  HST and  Chandra observations
were not taken simultaneously.  However, it is evident from the values
of Table 2  and the values of Tables  3 and 6 of Harris  et al. (2006)
that the rise  and fall timescales are consistently  larger in the NUV
than in the X-rays. See,  for instance, the time interval between 2005
June 21  and 2005 August 06 when  halving time for the  X-rays is 0.21
while in  the NUV between  2005 June 22  and 2005 August 01  the decay
time is 0.36.  Harris et al.  (2008) make a  detailed analysis of rise
and fall  time scales  for HST-1 and  conclude that this  longer decay
time in  the NUV is  an indication that  expansion is not  the primary
energy loss  mechanism for the  charged particles emitting  within the
HST-1 region.

The detection  of polarized emission  as well as  synchrotron emission
models fitted to flux measurements provided evidence that the physical
process responsible for  the radio to UV emission in  the knots of the
M87  jet is  synchrotron  radiation of  electrons  accelerated by  the
magnetic field  of the  jet (Perlman et  al.  2001).  The  very strong
correlation between the NUV and  X-ray light curves of the HST-1 flare
proves that  the same physical  phenomenon and the same  electrons are
responsible  for the  emission  in both  bands.   Therefore the  X-ray
emission  is  also  synchrotron  in  origin. The  injection  of  fresh
particles  into  the flaring  volume  is  not  needed to  explain  the
high-energy emission,  the X-ray emission  is well interpreted  as the
high energy  extension of  the radio to  optical spectrum  (Perlman \&
Wilson 2005,  Harris et al.   2006).  The observations  presented here
rules  out   inverse  Compton  (IC)  up-scattering   of  lower  energy
electrons, as the  cause of the high energy emission  of this flare in
particular,  and  of the  other  emission  knots  along the  M87  jet.
Moreover, high-energy photons produced  by IC up-scattering would take
much longer (10000  years) than the observed time  to decrease in flux
(Harris, 2003).We  can thus safely state that  synchrotron emission is
the physical process  responsible for the high energy  emission in the
M87 jet.

The encircled energy distribution of  HST-1 follows the pattern of the
detectors PSF at all epochs. In the NUV, HST-1 remains unresolved with
an  upper limit  on  its size  of  $\sim 0.025\arcsec$,  i. e.   $\sim
1.9pc$.

The HESS  collaboration recently reported a detection  of TeV $\gamma$
rays emanating  from M87, but  the Cherenkov telescopes used  for this
detection lack the  resolution to determine the exact  position of the
TeV emitting region. Cheung et al. (2007) favor HST-1 over the nucleus
as the  site of origin of the  TeV emission. They note  that the light
curve of TeV emission in M87  is roughly similar to the light curve of
the  HST-1  flare  seen  in  the  radio  and  X-ray.   The  NUV  peaks
simultaneously with  the X-rays and  the $\gamma$ rays and  only HST-1
shows  a flaring  behavior in  the NUV.   The nucleus  shows  only the
characteristic low  amplitude variability, see Figure  2.  These facts
support the hypothesis of HST-1 as  the site of origin of the $\gamma$
rays  through  IC  upscattering  of  ambient  photons  by  high-energy
electrons produced during the outburst.

After  2003 May,  and for  more  than four  years, the  flux of  HST-1
dominates the NUV emission  of M87, patently overpowering the emission
of the central  engine (see Fig. 3). Given  that within radio galaxies
the principal  sources of particle  acceleration are the core  and the
jet, the  large flux from this flare  plays, as we have  shown here, a
crucial role  determining the NUV  flux and therefore the  spectrum of
the entire galaxy.

As part  of the  Chandra Cen  A Very Large  Project Hardcastle  et al.
(2007) searched, to no avail,  for HST-1-like variability in the X-ray
jet of this galaxy (D=3.7Mpc).   Hardcastle et al.  aimed at answering
an important question:  Is HST-1 a feature unique of  M87?  Or is this
extreme variability ubiquitous, or at least frequent, in AGN jets?  If
an HST-1-type outburst  occurred in more distant AGN,  it would not be
resolved with current optical and X-ray instruments and would probably
be associated with  Doppler boosting of emission emanating  from a jet
close to the line of sight or with events associated to variability of
the accretion disk  of the black hole.  However, the  angle of the M87
jet with  the line of sight is  about 26-30 deg (Cheung  et al.  2007,
Bicknell  \& Begelman  1996) allowing  only a  modest  beaming.  Also,
given its large  distance from the core, i.e.  more than  65 pc or one
million  Schwarzchild radii,  intrinsic black  hole variability  has no
direct relation with this flaring.   Outbursts similar to HST-1 can be
responsible for variability associated  with high redshift blazars but
remain completely unresolved.


\acknowledgments

This  research has  made  use  of the  NASA  Astrophysics Data  System
Bibliographic  services.  I  wish  to thank  Jennifer  Mack and  Marco
Sirianni (STScI) for answering an  endless list of questions about the
ACS. I  am grateful to Ethan  Vishniac (McMaster) for  believing in my
understanding of  accretion disks. The  anonymous referee gave  a very
constructive report that helped  to improve this paper. Laura Schwartz
(JHU) encouraged me to carry out this project and many others.


\begin{figure}
\plotone{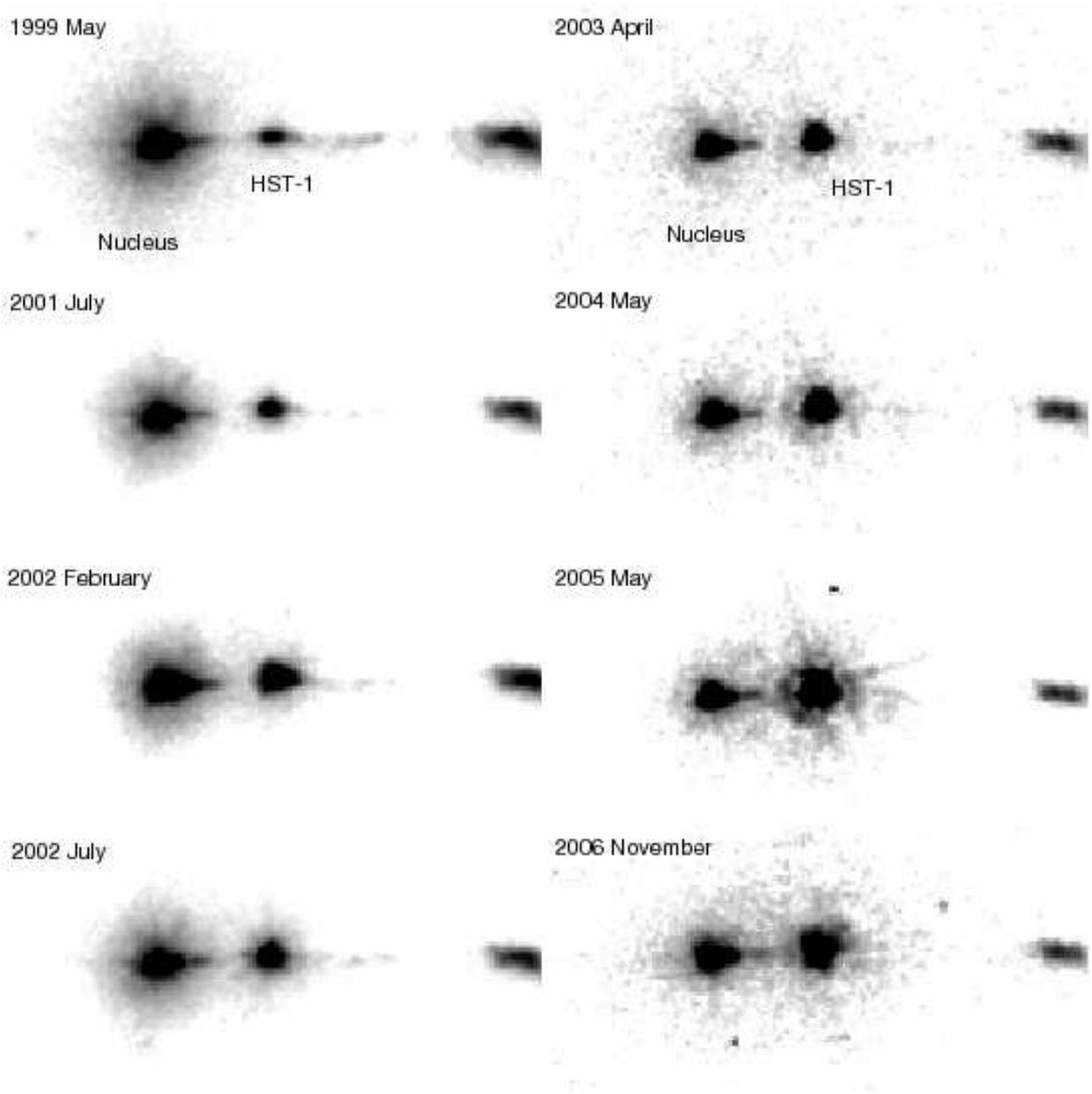}

\caption{Observations  of  the  inner   jet  of  M87  taken  with  the
  STIS/F25QTZ  (left column)  and the  ACS/F220W (right  column).  The
  nucleus is on  the left, HST-1 is the variable  feature to the right
  as noted on  the top image of each column.   The observation date is
  on the upper  left hand corner of each image.   The total jet length
  shown in these images is about $3\arcsec$ or 230 parsecs. The jet
  has been rotated to be aligned with the horizontal axis.  }

\end{figure}


\begin{figure}
\plotone{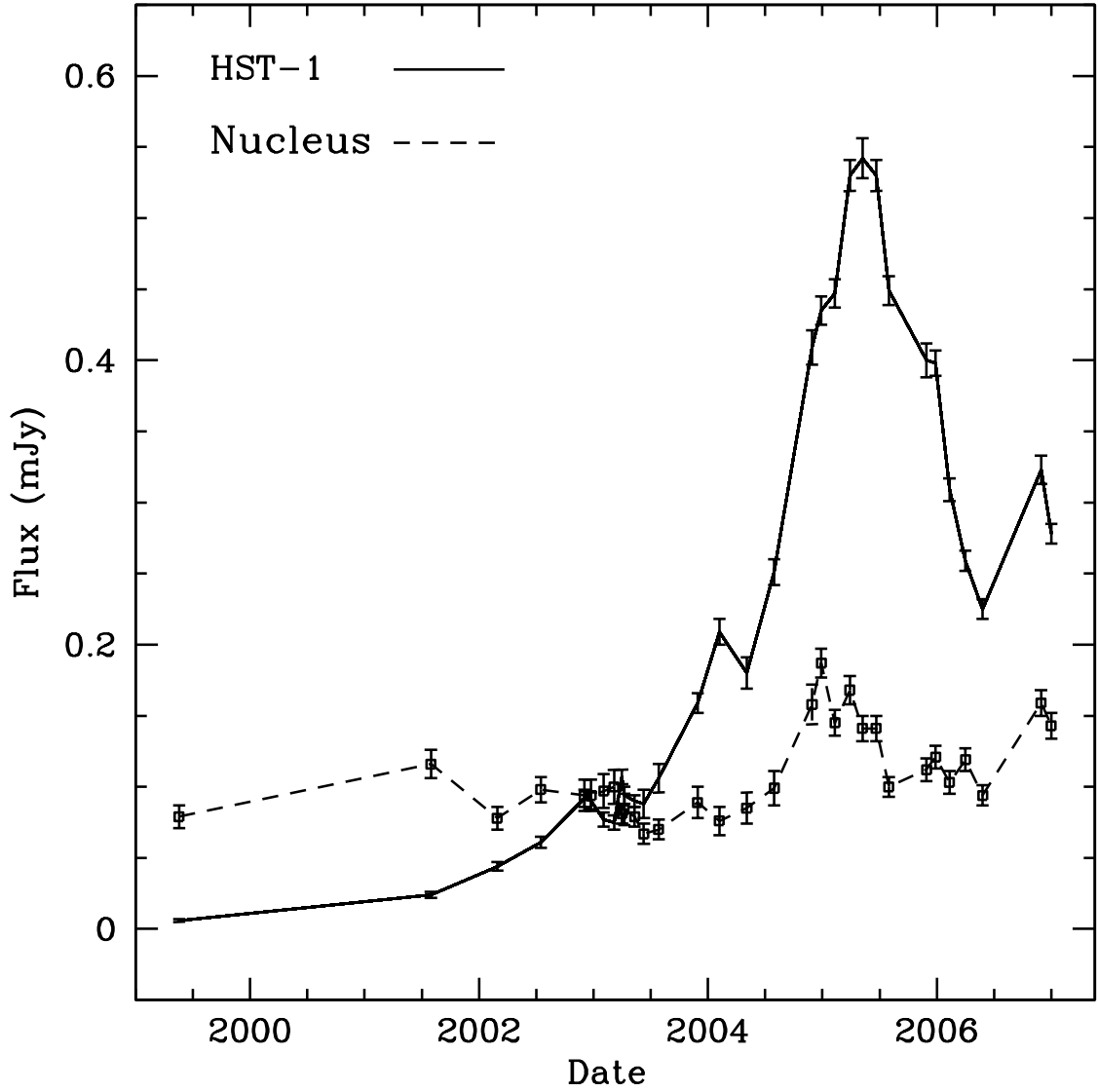}
\caption{NUV light curve for knot HST-1 and the nucleus of M87 from May
  1999  to December 2006,  in units  of milliJanskys  (mJy), $1Jy=10^{-26}W
  Hz^{-1} m^{-2}$.}
\end{figure}


\begin{figure}
\plotone{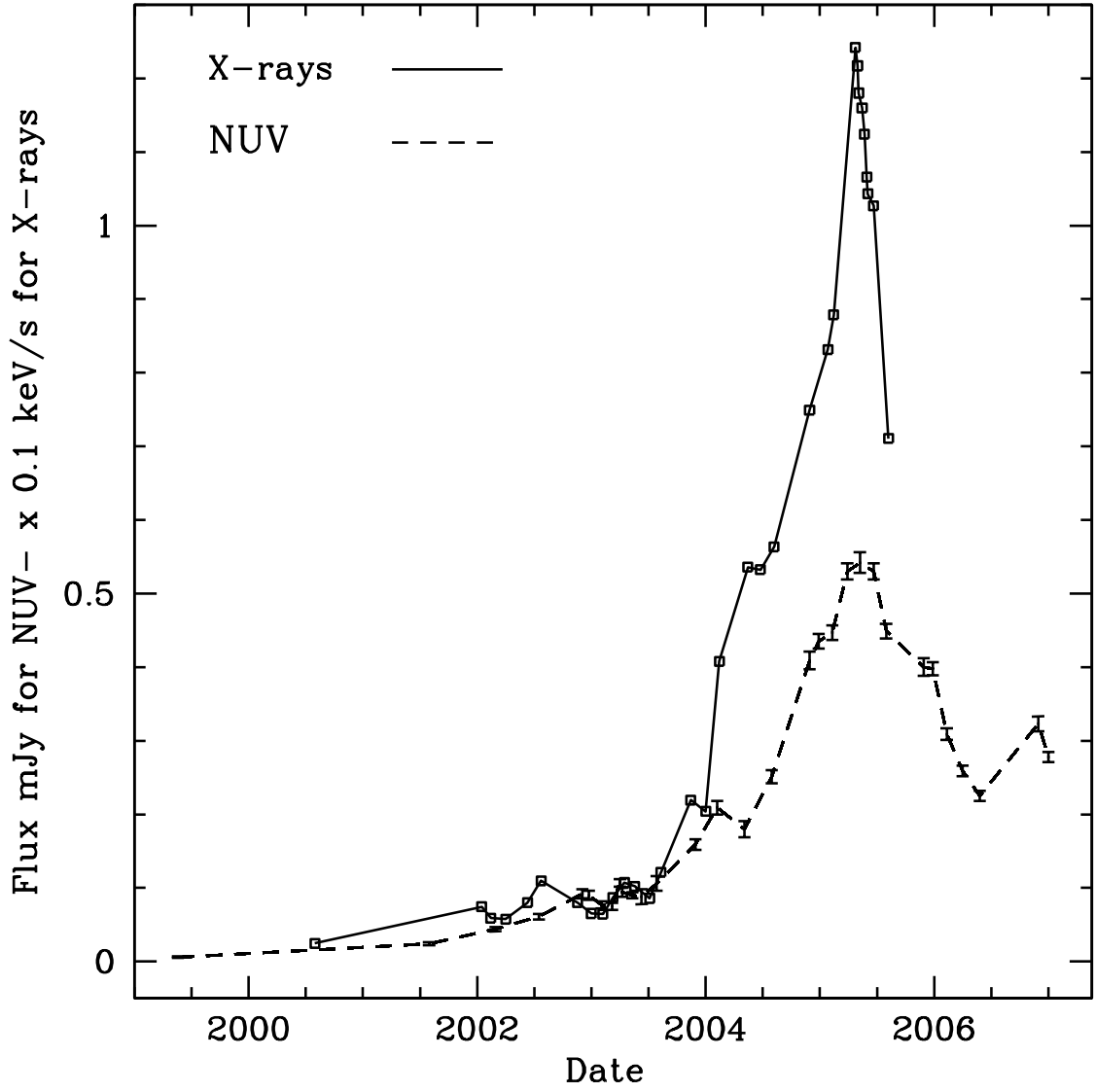}
\caption{NUV and X-ray light curves of HST-1. The X-ray data was taken
  from Harris et al. (2006).}
\end{figure}



\begin{deluxetable}{llccccc}
\tabletypesize{\small}
\tablecaption{Log of observations  and NUV intensities of the Nucleus
  and HST-1\label{tbl-2}} 
\tablewidth{0pt} \tablehead{\colhead{Date} & \colhead{Camera/Filter} & \colhead{Nucleus}
& \colhead{HST-1} }
 
\startdata

1999 May 17  &  STIS/F25QTZ     & 0.079   $\pm$ 0.008 &   0.006 $\pm$ 0.001 \\
2001 Jul 30  &  STIS/F25QTZ     & 0.116   $\pm$ 0.010 &   0.024 $\pm$ 0.002 \\
2002 Feb 27  &  STIS/F25QTZ     & 0.078   $\pm$ 0.008 &   0.044 $\pm$ 0.003 \\
2002 Jul 17  &  STIS/F25QTZ     & 0.098   $\pm$ 0.009 &   0.061 $\pm$ 0.004 \\
2002 Nov 30  &  ACS/F220W       & 0.094   $\pm$ 0.011 &   0.092 $\pm$ 0.006 \\
2002 Dec 22  &  ACS/F220W       & 0.094   $\pm$ 0.011 &   0.090 $\pm$ 0.006 \\
2003 Feb 02  &  ACS/F220W       & 0.097   $\pm$ 0.012 &   0.077 $\pm$ 0.005 \\
2003 Mar 06  &  ACS/F220W       & 0.100   $\pm$ 0.012 &   0.075 $\pm$ 0.005 \\
2003 Mar 31  &  ACS/F250W       & 0.081   $\pm$ 0.007 &   0.107 $\pm$ 0.005 \\
2003 Apr 07  &  ACS/F220W       & 0.084   $\pm$ 0.011 &   0.094 $\pm$ 0.006 \\ 
2003 May 10  &  ACS/F250W       & 0.079   $\pm$ 0.007 &   0.090 $\pm$ 0.004 \\ 
2003 Jun 7   &  STIS/F25QTZ     & 0.067   $\pm$ 0.007 &   0.088 $\pm$ 0.010 \\
2003 Jul 27  &  STIS/F25QTZ     & 0.070   $\pm$ 0.007 &   0.106 $\pm$ 0.010 \\
2003 Nov 29  &  ACS/F220W       & 0.089   $\pm$ 0.011 &   0.159 $\pm$ 0.007 \\
2004 Feb 07  &  ACS/F220W       & 0.076   $\pm$ 0.010 &   0.209 $\pm$ 0.009 \\
2004 May 05  &  ACS/F220W       & 0.085   $\pm$ 0.011 &   0.180 $\pm$ 0.011 \\ 
2004 Jul 30  &  ACS/F220W       & 0.099   $\pm$ 0.012 &   0.251 $\pm$ 0.009 \\
2004 Nov 28  &  ACS/F220W/F250W & 0.158   $\pm$ 0.014 &   0.409 $\pm$ 0.012 \\       
2004 Dec 26  &  ACS/F250W       & 0.187   $\pm$ 0.010 &   0.435 $\pm$ 0.010 \\
2005 Feb 09  &  ACS/F250W       & 0.145   $\pm$ 0.009 &   0.447 $\pm$ 0.010 \\
2005 Mar 27  &  ACS/F250W       & 0.168   $\pm$ 0.010 &   0.530 $\pm$ 0.011 \\
2005 May 09  &  ACS/F220W/F250W & 0.141   $\pm$ 0.009 &   0.542 $\pm$ 0.014 \\
2005 Jun 22  &  ACS/F250W       & 0.141   $\pm$ 0.009 &   0.530 $\pm$ 0.011 \\
2005 Aug 01  &  ACS/F250W       & 0.100   $\pm$ 0.007 &   0.449 $\pm$ 0.010 \\
2005 Nov 29  &  ACS/F220W/F250W & 0.112   $\pm$ 0.008 &   0.400 $\pm$ 0.012 \\
2005 Dec 26  &  ACS/F250W       & 0.121   $\pm$ 0.008 &   0.398 $\pm$ 0.009 \\
2006 Feb 08  &  ACS/F220W/F250W & 0.103   $\pm$ 0.008 &   0.309 $\pm$ 0.008 \\
2006 Mar 30  &  ACS/F250W       & 0.119   $\pm$ 0.008 &   0.259 $\pm$ 0.007 \\
2006 May 23  &  ACS/F220W/F250W & 0.094   $\pm$ 0.007 &   0.225 $\pm$ 0.007 \\
2006 Nov 28  &  ACS/F220W/F250W & 0.159   $\pm$ 0.009 &   0.323 $\pm$ 0.010 \\
2006 Dec 30  &  ACS/F250W       & 0.143   $\pm$ 0.009 &   0.278 $\pm$ 0.007 \\

 \enddata

 \tablecomments{Col.   (1), observation dates;  col.  (2),  camera and
   filter  in  use;  col  (3),  nucleus  flux  and  poisson  error  in
   milliJanskys;  col.   (4), flux  and  poisson  error  for HST-1  in
   milliJanskys. }

\end{deluxetable}


\begin{deluxetable}{lcccc}

  \tabletypesize{\small} \tablecaption{NUV  doubling and halving times
    (DT    and    HT)    for   HST-1\label{tbl-2}}    \tablewidth{0pt}
  \tablehead{\colhead{Epoch   Interval}    &   \colhead{$\Delta$t}   &
    \colhead{y-1} & \colhead{DT}& \colhead{HT} }
 
\startdata

1999 May 17 - 2001 Jul 30  & 2.21  & 3.00   & 0.73 &     \\
2001 Jul 30 - 2002 Feb 27  & 0.58  & 0.83   & 0.70 &     \\
2002 Feb 27 - 2002 Jul 17  & 0.38  & 0.39   & 0.97 &     \\
2002 Jul 17 - 2002 Nov 30  & 0.37  & 0.51   & 0.72 &     \\
2002 Nov 30 - 2002 Dec 22  & 0.06  & 0.02   &      & 1.50\\
2002 Dec 22 - 2003 Feb 02  & 0.12  & 0.14   &      & 0.43\\
2003 Feb 02 - 2003 Mar 06  & 0.09  & 0.03   &      & 1.50\\
2003 Mar 06 - 2003 Mar 31  & 0.07  & 0.43   & 0.16 &     \\
2003 Mar 31 - 2003 Apr 07  & 0.02  & 0.12   &      & 0.08\\
2003 Apr 07 - 2003 May 10  & 0.09  & 0.04   &      & 1.13\\ 
2003 May 10 - 2003 Jun 07  & 0.08  & 0.22   &      & 0.18\\
2003 Jun 07 - 2003 Jul 27  & 0.14  & 0.21   & 0.67 &     \\
2003 Jul 27 - 2003 Nov 29  & 0.34  & 0.50   & 0.68 &     \\
2003 Nov 29 - 2004 Feb 07  & 0.19  & 0.31   & 0.61 &     \\
2004 Feb 07 - 2004 May 05  & 0.24  & 0.14   &      & 0.85\\
2004 May 05 - 2004 Jul 30  & 0.24  & 0.39   & 0.61 &     \\
2004 Jul 30 - 2004 Nov 28  & 0.33  & 0.63   & 0.52 &     \\
2004 Nov 28 - 2004 Dec 26  & 0.08  & 0.06   & 1.33 &     \\
2004 Dec 26 - 2005 Feb 09  & 0.12  & 0.03   & 4.44 &     \\
2005 Feb 09 - 2005 Mar 27  & 0.12  & 0.19   & 0.63 &     \\
2005 Mar 27 - 2005 May 09  & 0.12  & 0.02   & 5.22 &     \\
2005 May 09 - 2005 Jun 22  & 0.12  & 0.02   &      & 5.22\\
2005 Jun 22 - 2005 Aug 01  & 0.11  & 0.15   &      & 0.36\\
2005 Aug 01 - 2005 Nov 29  & 0.33  & 0.11   &      & 1.50\\
2005 Nov 29 - 2005 Dec 26  & 0.07  & 0.01   &      & 3.50\\
2005 Dec 26 - 2006 Feb 08  & 0.12  & 0.22   &      & 0.27\\
2006 Feb 08 - 2006 Mar 30  & 0.14  & 0.16   &      & 0.44\\
2006 Mar 30 - 2006 May 23  & 0.15  & 0.13   &      & 0.57\\
2006 May 23 - 2006 Nov 28  & 0.52  & 0.44   & 1.18 &     \\
2006 Nov 28 - 2006 Dec 30  & 0.09  & 0.14   &      & 0.32\\

 \enddata

 \tablecomments{Col.  (1),  dates; col.  (2),  $\Delta$t time interval
   in years ; col (3),  y-1 where y=$I_{2}/I_{1}$, 1-y if $I_{2}<I_{1}$
   ; col.  (4), doubling time (DT) or halving time (HT). }

\end{deluxetable}

\clearpage

\end{document}